\documentclass[aps,prl,preprint,tightenlines,superscriptaddress,showpacs,byrevte
x]{revtex4}

\usepackage{graphicx}
\usepackage{bm}
\usepackage{xspace}
\usepackage{amsmath}
\usepackage{amssymb}

\newcommand{\ABS}[1]{\mid\!#1\!\mid}
\def\BB{\ensuremath{B\overline{B}}\xspace}
\def\Br{\ensuremath{{\cal{B}}}}
\def\dd{\ensuremath{D^0\overline{D}{}^{0}}\xspace}
\def\ddp{\ensuremath{D^0\overline{D}{}^{0}\pi^0}\xspace}
\def\ddst{\ensuremath{D^0\overline{D}{}^{*0}}\xspace}
\def\de{\ensuremath{\Delta E}\xspace}
\def\ks {\ensuremath{K^0_{S}}\xspace}
\def\mbc{\ensuremath{M_\mathrm{bc}}\xspace}
\def\ttr{\ensuremath{\theta_{\rm Thrust}}\xspace}
\def\to {\ensuremath{\rightarrow}}
\def\bddstk{\ensuremath{B \to D^0\overline{D}{}^{*0} K}\xspace}
\def\bddpk{\ensuremath{B \to D^0\overline{D}{}^{0}\pi^0 K}\xspace}

\begin{document}

\title{\Large\bf Observation of a Near-threshold $\ddp$
Enhancement in $B \to \ddp K$ Decay }

\affiliation{Budker Institute of Nuclear Physics, Novosibirsk}
\affiliation{Chiba University, Chiba}
\affiliation{University of Cincinnati, Cincinnati, Ohio 45221}
\affiliation{Gyeongsang National University, Chinju}
\affiliation{University of Hawaii, Honolulu, Hawaii 96822}
\affiliation{High Energy Accelerator Research Organization (KEK), Tsukuba}
\affiliation{Hiroshima Institute of Technology, Hiroshima}
\affiliation{University of Illinois at Urbana-Champaign, Urbana, Illinois 61801}
\affiliation{Institute of High Energy Physics, Vienna}
\affiliation{Institute of High Energy Physics, Protvino}
\affiliation{Institute for Theoretical and Experimental Physics, Moscow}
\affiliation{J. Stefan Institute, Ljubljana}
\affiliation{Kanagawa University, Yokohama}
\affiliation{Korea University, Seoul}
\affiliation{Swiss Federal Institute of Technology of Lausanne, EPFL, Lausanne}
\affiliation{University of Ljubljana, Ljubljana}
\affiliation{University of Maribor, Maribor}
\affiliation{University of Melbourne, Victoria}
\affiliation{Nagoya University, Nagoya}
\affiliation{Nara Women's University, Nara}
\affiliation{National Central University, Chung-li}
\affiliation{National United University, Miao Li}
\affiliation{Department of Physics, National Taiwan University, Taipei}
\affiliation{H. Niewodniczanski Institute of Nuclear Physics, Krakow}
\affiliation{Nippon Dental University, Niigata}
\affiliation{Niigata University, Niigata}
\affiliation{University of Nova Gorica, Nova Gorica}
\affiliation{Osaka City University, Osaka}
\affiliation{Osaka University, Osaka}
\affiliation{Panjab University, Chandigarh}
\affiliation{Peking University, Beijing}
\affiliation{Princeton University, Princeton, New Jersey 08544}
\affiliation{RIKEN BNL Research Center, Upton, New York 11973}
\affiliation{University of Science and Technology of China, Hefei}
\affiliation{Sungkyunkwan University, Suwon}
\affiliation{University of Sydney, Sydney NSW}
\affiliation{Tata Institute of Fundamental Research, Bombay}
\affiliation{Toho University, Funabashi}
\affiliation{Tohoku Gakuin University, Tagajo}
\affiliation{Tohoku University, Sendai}
\affiliation{Department of Physics, University of Tokyo, Tokyo}
\affiliation{Tokyo Institute of Technology, Tokyo}
\affiliation{Tokyo Metropolitan University, Tokyo}
\affiliation{Tokyo University of Agriculture and Technology, Tokyo}
\affiliation{Virginia Polytechnic Institute and State University, Blacksburg, Virginia 24061}
\affiliation{Yonsei University, Seoul}
   \author{G.~Gokhroo}\affiliation{Tata Institute of Fundamental Research, Bombay} 
   \author{G.~Majumder}\affiliation{Tata Institute of Fundamental Research, Bombay} 
   \author{K.~Abe}\affiliation{High Energy Accelerator Research Organization (KEK), Tsukuba} 
   \author{K.~Abe}\affiliation{Tohoku Gakuin University, Tagajo} 
   \author{I.~Adachi}\affiliation{High Energy Accelerator Research Organization (KEK), Tsukuba} 
   \author{H.~Aihara}\affiliation{Department of Physics, University of Tokyo, Tokyo} 
   \author{D.~Anipko}\affiliation{Budker Institute of Nuclear Physics, Novosibirsk} 
   \author{T.~Aushev}\affiliation{Institute for Theoretical and Experimental Physics, Moscow} 
   \author{T.~Aziz}\affiliation{Tata Institute of Fundamental Research, Bombay} 
   \author{A.~M.~Bakich}\affiliation{University of Sydney, Sydney NSW} 
% ...
\collaboration{The Belle Collaboration}

\begin{abstract}
 
We report the first observation of a near-threshold  enhancement in the 
$\ddp$ system from $\bddpk$ decays using a 414 $\mathrm{fb}^{-1}$ data sample 
collected  at the $\Upsilon(4S)$ resonance. The
enhancement peaks at a mass  $M=3875.4 \pm 0.7_{-2.0}^{+1.2}$~MeV/$c^2$ and
the branching fraction for events in the peak is 
\Br ($B \to \ddp K$)=(1.27 $\pm$ 0.31 $_{-0.39}^{+0.22})$ $\times$ $10^{-4}$.
The data were collected with the Belle detector at the KEKB
energy-asymmetric $e^+e^-$ collider.

\end{abstract}

\maketitle

Belle recently discovered~\cite{xdiscover} a new state, the $X(3872)$, with a
mass of (3872.0 $\pm$ 0.6 $\pm$ 0.5) MeV/$c^2$ and width less than 2.3 MeV/$c^2$ in
the $J/\psi\pi^+\pi^-$ system from $B^\pm \to J/\psi\pi^+\pi^- K^\pm$ decays.
Other experiments confirmed the existence of the X(3872) \cite{x3872oth},
but all published results are in the $J/\psi \pi^+ \pi^-$ mode only.
Although the initial expectations were that the $X(3872)$ was one
of the unobserved charmonium states, subsequent experimental 
observations disfavor this hypothesis~\cite{properties}.
 
Since the properties of the $X(3872) $
are not consistent with a charmonium assignment, there
have been speculations that it is some type of exotic
state, for example, a $q\overline{q}g$ hybrid ~\cite{hybrid}.
The $X(3872)$ mass is within errors of the $\ddst $ threshold 
($3871.2\pm 0.9$~MeV/$c^2$), which triggered speculation that 
it might be a $\ddst$ bound state (deuson)~\cite{deuson_ref}. If 
the $X(3872)$ is a  loosely bound S-wave molecule composed of $\ddst$ charm 
mesons, it is expected that there will be an enhancement in the 
near-threshold $\ddst$ invariant mass 
distribution~\cite{deuson_ref,braaten}. The Belle collaboration 
found no evidence for $D^0\overline{D}{}^0$, $D^+D^-$ and 
$D^0\overline{D}{}^0\pi^0$ decays of $X(3872)$ with a smaller sample of 
$\BB$ events \cite{nullres}. 
The distributions of daughter particle momenta for $\ddp$ decays
of a $\ddst$ molecule are expected to be different from those of an
incoherent sum of the decays of free $D^{*0}$ and 
$\overline{D}{}^0 $~\cite{voloshin}.

In this letter, the $\ddp$ system is studied in 
$B^+ \to \ddp K^+$ and $B^0 \to \ddp \ks$ decays.
Inclusion of charge conjugate modes is implied throughout this paper.

These results are
based on a 414 $\mathrm{fb}^{-1}$ data sample corresponding to 447 million 
$B\overline{B}$ 
pair events collected with the Belle detector \cite{belledet} at the 
energy-asymmetric $e^+e^-$ collider KEKB \cite{kekb}. 
The fractions of neutral and charged $B$ mesons produced in the
$\Upsilon(4S)$ peak are assumed to be equal.
 
The Belle detector is a general purpose magnetic spectrometer with a 1.5~T 
magnetic field provided by a superconducting solenoid. 
Momenta of charged particles are measured using a silicon
vertex detector and a 50-layer central drift chamber(CDC). Photons are 
detected in an electromagnetic calorimeter (ECL) consisting of 8736 
CsI(Tl) crystals. 
Particle identification likelihoods, ${\mathcal{L}}_{\pi/K}$, are 
derived from
the information provided by an array of 128
time-of-flight counters, an array of 1188 silica aerogel 
\v{C}erenkov threshold counters and $dE/dx$-measurements in the CDC.

Kaon candidates are selected from well-measured tracks by using a 
requirement on the likelihood 
ratio, ${\mathcal{L}}_K/({\mathcal{L}}_K+{\mathcal{L}}_\pi)$, which has an average kaon identification efficiency $\sim$ 97\% with
a pion misidentification rate of $\sim$ 18\%.
Similarly, charged pions are selected with an efficiency of $\sim$ 98\%
and kaon misidentification rate of $\sim$ 12\%. 
All tracks 
compatible with the electron hypothesis ($\sim$ 0.2\% 
misidentification rates from pion/kaon) are eliminated. 

Neutral pions ($\pi^0$) are reconstructed from pairs of isolated ECL clusters 
(photons) with invariant mass in the range 119 MeV/$c^2$ $< M_{\gamma \gamma} <$ 
150 MeV/$c^2$ ($\sim \pm 3 \sigma$). The energy of each photon is required to 
be greater than 30 MeV in the barrel region, defined as $32^\circ < 
\theta_\gamma < 128^\circ$, and greater than 50 MeV in the endcap regions, 
defined as $17^\circ < \theta_\gamma \leq 32^\circ$ or $128^\circ < 
\theta_\gamma \leq 150^\circ$, where $\theta_\gamma$ denotes the polar angle 
of the photon. A mass-constrained fit is applied to obtain the four-momentum 
of a $\pi^0$ candidate.

Neutral kaons ($\ks$) are reconstructed {\it via} the $\ks \to \pi^+ \pi^-$ 
decay mode.  There is no particle identification requirement for daughter pions and 
the requirement  \mbox{$\ABS{M_{\pi\pi} - M_{ \ks }} < 11 $ MeV/$c^2$}
($\sim 3.5 \sigma$, where $\sigma$ is the $\pi^+ \pi^-$ invariant mass
resolution) is applied. Selection criteria to reduce random 
combinations of two tracks are described in detail elsewhere \cite{ksref}.
A mass-vertex
constrained fit is performed to the $\ks$ candidate to improve the 
resolution on its momentum measurement.
 
Beam-gas events are rejected using the requirements $|P_z| <$ 2\,GeV/$c$
and 1.0 $< E_{\rm vis}/E_{\rm beam}<$ 2.5, where $P_z$, $E_{\rm beam}$ and 
$E_{\rm vis}$ are the longitudinal momentum sum, beam energy and total visible 
energy, respectively, in the center of mass (CM) frame.  Continuum
events ($e^+e^- \to q\overline{q}$, where $q=u,d,s,c$)  are suppressed
by requirements on the ratio of the second to the zeroth Fox-Wolfram 
moments~\cite{foxwol}, $R_2 <$ 0.50.

Candidate ${D}^0$ mesons are reconstructed from well-measured charged tracks
in the $K^-\pi^+$, $K^-\pi^+\pi^+\pi^-$, $\ks \pi^+ \pi^-$ and
$K^+ K^-$ decay modes. A $\pm$3 sigma mass window is applied for 
selecting $D^0$'s, where sigma is the decay-mode-dependent resolution 
of the reconstructed $D^0$ mass (typically $\sim$ 4.5 MeV/$c^2$).
Mass and vertex 
constrained fits are applied to improve the $D^0$ meson momentum 
resolution. 

A $\dd$ candidate pair is combined with a $\pi^0$ and a kaon to 
reconstruct a $B$ candidate. Continuum events are further suppressed 
with the
criterion, $|\cos\ttr|< 0.9$, where $\ttr$ is the angle between the 
thrust axis of the $B$ candidates and the thrust axis of the remaining 
tracks and isolated ECL clusters. The beam-energy constrained
mass, $ \mbc$ ($= \sqrt{(E_{\rm beam})^2 - (\sum_i \vec{P}_i)^2}$), where 
$\vec{P}_i$ is the momentum of 
the $i$th daughter of the candidate $B$ in the CM frame is 
restricted between 5.273 GeV/$c^2$ and 5.286 GeV/$c^2$.
A peak in the difference between the measured energy of the $B$ candidate and 
the beam energy, $\de$ (=$ \sum_i E_i - E_{\rm beam}$) is
a signature of $B$-meson signal events, where $E_i$ is the CM energy of 
the $i$th daughter of the candidate $B$.

The $B \to \ddp K$ signal Monte Carlo (MC) sample is generated in two steps, 
$B \to X(3872)K$ followed by $X(3872) \to \ddp$ assuming a phase space 
distribution in both decay chains.
The average number of $\ddp K$ entries per MC signal event is $\sim$ 3.55, 
which are mainly due to multiple slow $\pi^0$'s. The characteristics 
of incorrectly reconstructed $\pi^0$ candidates are identical to signal $\pi^0$'s. 
To reduce this multiplicity we use the selection criterion,
$M_{D^0\pi^0} < 2.013$ GeV$/c^2$ or 
$M_{\overline{D}{}^0\pi^0} < 2.013$ GeV$/c^2$.
This requirement reduces background and the candidate multiplicity to 1.68
with almost no loss of signal efficiency. 
Possible bias due to this selection criterion is studied in the 
following event samples: 
$(i)$ a large sample of generic $B\overline{B}$ and continuum MC events;
$(ii)$ $D^0D^-\pi^0 K$ data;
$(iii)$ $D^0D^0\pi^0 K$ (same-flavor charm) data;
$(iv)$ the $D^0$-mass sideband data (one $D^0$-meson is reconstructed when the
invariant mass of daughters is outside the $D^0$-mass signal region); and
$(v)$ $\de$ side-band data (60 MeV $<|\de|<$ 110 MeV).
No peaking behavior is observed in the $DD\pi^0$ mass distribution for any 
of the above-mentioned control samples, thereby confirming that there is 
no bias in the selection criteria on $M_{\ddp}$.

A unique $\ddp K$ candidate is chosen out of possible multiple candidates
in a given event by taking the combination 
with the  minimum value of
   $\left({\Delta M_{\pi^0}\over\sigma_{ M_{\pi^0}}}\right)^2 +
    \left({\Delta M_{D^0}\over\sigma_{M_{D^0}}}\right)^2 +
    \left({\Delta M_{\overline{D}{}^0}\over
           \sigma_{M_{\overline{D}{}^0}}}\right)^2$,
where $\Delta x$ and 
$\sigma_{x}$ are the deviation of the measured quantity $x$ from its 
nominal value and the uncertainty 
in its measurement, respectively.  Multiple kaon entries are 
resolved  by choosing the candidate with the highest kaon identification 
probability for charged kaon and minimum $|M_{\pi\pi} - M_{\ks}|$
for neutral kaon. There is a negligibly small number of 
events with charged and neutral kaon multiple entries.  

An unbinned extended maximum likelihood fit to the $\ddp$ 
invariant mass, $M_{\ddp}$, and $\de$ distributions is
used to obtain the signal yield.  The fit includes three components:
$(i)$ a signal function, which is modelled by the
sum of two Gaussian functions with the same mean value for $\de$ 
and single Gaussian function
for $M_{\ddp}$; $(ii)$ a non-resonant $B \to \ddst K$ signal, where
$\de$ is also modelled by a double Gaussian function and $M_{\ddp}$ 
with a threshold function; and 
$(iii)$ the remaining backgrounds, which are modelled with
a first-order polynomial for $\de$ and another threshold function
for $M_{\ddp}$, this threshold function is obtained from the
$\Delta E$ sideband data of $B \to \ddp K$ events. 
Shapes of the $\de$ distributions for signal and 
non-resonant $B \to \ddst K$ background are fixed from the
$B \to \ddst K$ data sample.   
The signal has a narrow Gaussian component with width 
$\sigma \sim 4.5$ MeV and a wide Gaussian component with width 4.6
times larger that accounts for 40\% of the signal.

Parameters of  the $M_{\ddp}$ threshold functions are fixed from 
a large MC sample of $B \to \ddst K$ events for the non-resonant 
components and  $B \to \ddp K$ $\de$ sideband data for remaining
backgrounds.  The normalization factor for the non-resonant
component is fixed according to the branching fraction from 
~\cite{bbarbtocck}.
The slope of the background polynomial, the parameters of $M_{\ddp}$ 
threshold peak, and the normalization
factors of signal and combinatorial background component are
free parameters of the fit.

\begin{figure}[htbp]
\begin{center}
\includegraphics[width=0.7\textwidth]{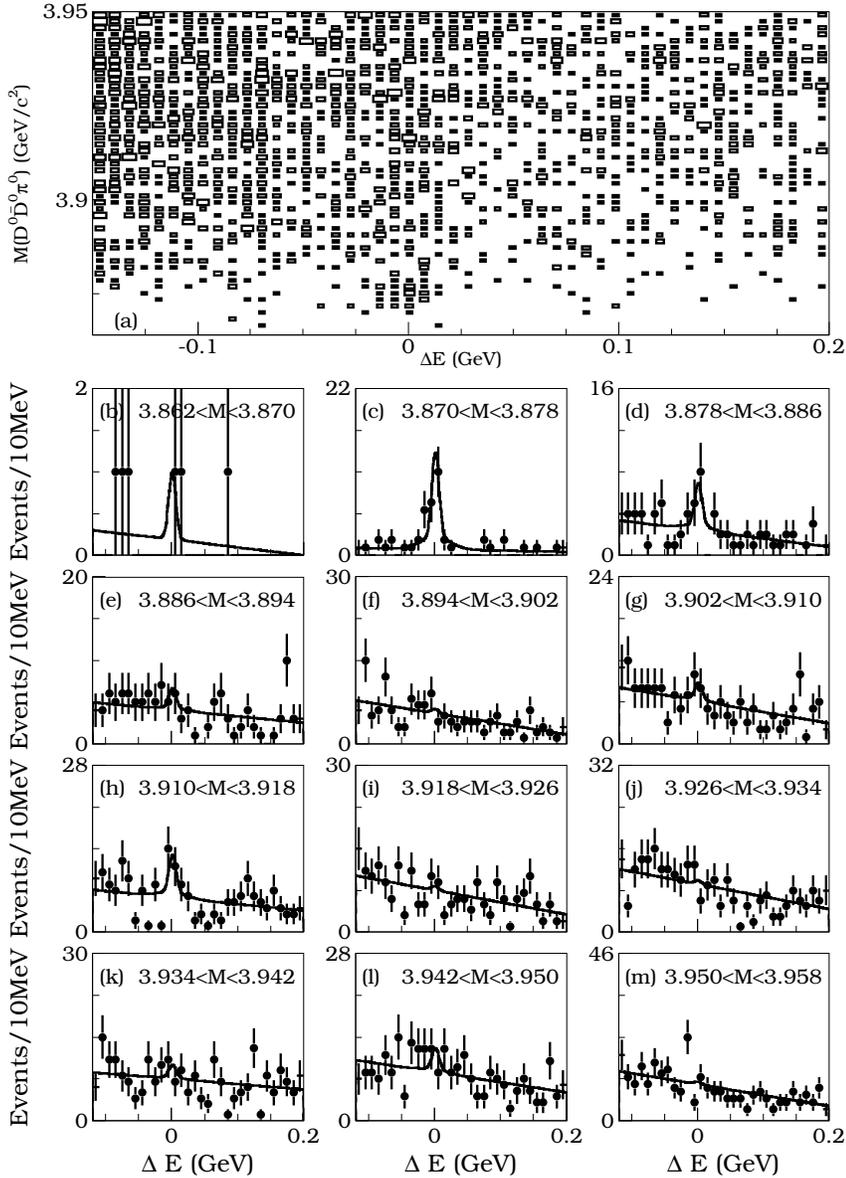}

\caption{(a) Scatter plot of $\de$ and $M_{\ddp}$ distribution in the data sample.
         (b)-(m) $\de$ distributions in 8 MeV/$c^2$ $\ddp$ mass 
         bins for a possible $B \to \ddp K $ signal in data.}
\label{fig:data_opt35_de8mv}
\end{center}
\end{figure}

Figure \ref{fig:data_opt35_de8mv}(a) shows the scatterplot of $\de$ and $M_{\ddp}$
in data. There is a cluster of events in the $\ddp$ threshold region.
The $\de$ distributions 
for different $M_{\ddp}$ intervals 
are shown in plots (b)-(m), where a one-dimensional fit gives a 
signal of (23.4$\pm$5.6) events in the $M_{\ddp}$ range from 
3.870 GeV/$c^2$ to 3.878 GeV/$c^2$.
The statistical significance of this signal, 
defined as $\sqrt{-2\,\ln\,({\cal{L}}_0/{\cal{L}}_{\max})}$, 
where ${\cal{L}}_{0(\max)}$ is the 
likelihood without (with) the signal contribution, is 6.4$\sigma$. 
A similar analysis that uses the $\mbc$ distribution rather than
the $\de$ distribution to measure the signal
also shows a clear peak for the same $M_{\ddp}$ interval, with a consistent
signal yield and similar statistical significance. 

\begin{figure}[htbp]
\begin{center}
\includegraphics[width=0.8\textwidth]{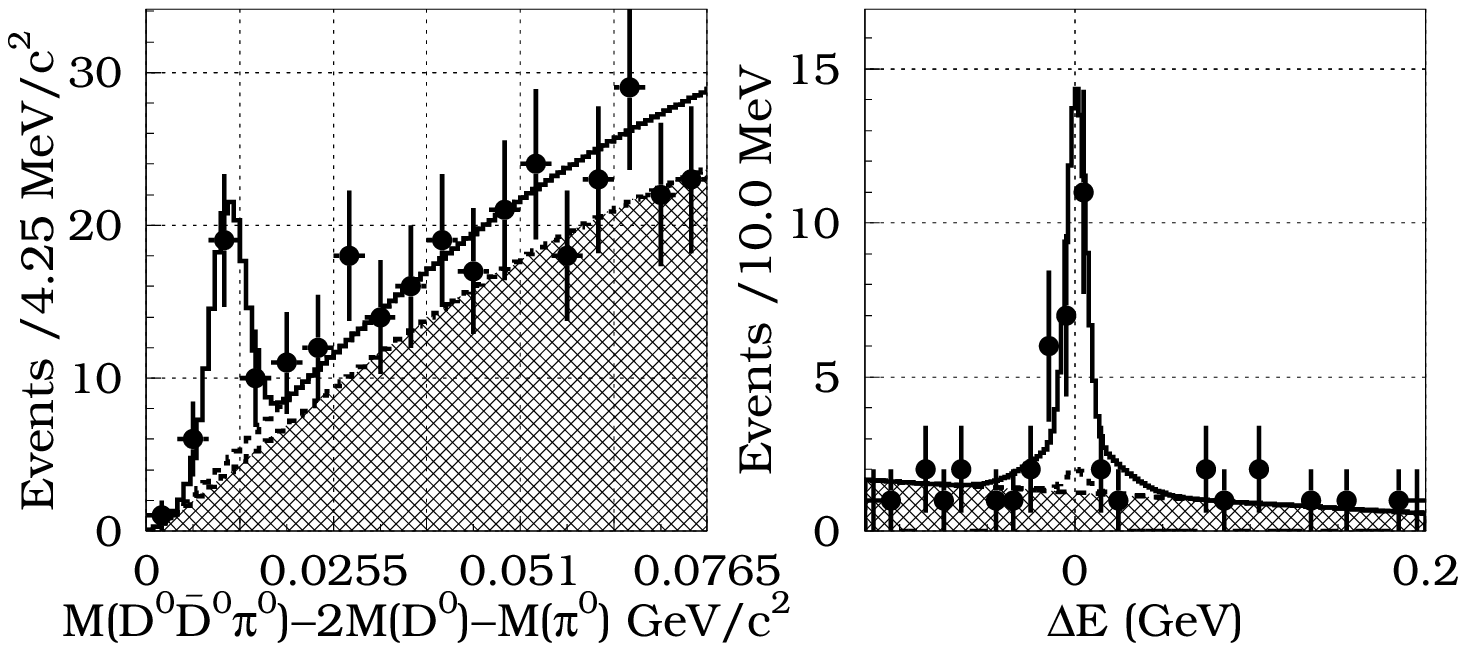}

\caption{Projection of the unbinned fit 
        (a) $Q$-value (=$M_{\ddp}-2M_{D^0}-M_{\pi^0}$) in
        $\Delta E$ signal region ($|\de |$ $<$ 25 MeV)
        and (b) $\de$ in the signal region of 
        $Q$-value (6 MeV/$c^2$ $<Q$-value $<$ 14 MeV/$c^2$).
        The dots are data points, the hatched histogram corresponds to
        combinatorial background; the dashed line indicates the total background 
        and the solid line is from the combined fitting function.}
\label{fig:2d_nonres_fix}
\end{center}
\end{figure}

To obtain the exact position of the near-threshold peak as well 
as its branching fraction, the two-dimensional distribution of $\de$ 
and $Q$-value (=$M_{\ddp}-2M_{D^0}-M_{\pi^0}$) is fitted.
Projections onto $Q$-value (for $|\de| <$ 25 MeV) 
and $\de$ (for 6 MeV/$c^2$ $<Q$-value $<$ 14 MeV/$c^2$) are 
shown in Figure~\ref{fig:2d_nonres_fix} along with the results 
of the fit.  The fitted mean and width of the near-threshold peak are 
11.21$\pm$0.68 MeV/$c^2$ and  2.42$\pm$0.55 MeV/$c^2$, respectively,
in the $Q$-value.
The signal yield is 24.1$\pm$6.1 and the significance,
including the effects of systematic errors, is 6.4$\sigma$.
Individual results for the charged and neutral 
$B$ meson samples are given in Table \ref{tab:bresult} together
with the combined result.

In terms of the invariant mass of $\ddp$ system, the peak position is
$M_{\ddp}$ = 3875.4 $\pm$0.7 MeV, where the error is statistical only.

{\tiny
\begin{table}[htbp]
\begin{center}
\caption{Signal yield and branching fraction for $B \to\ddp K$ with near-threshold 
         peak near 3.8754 GeV/$c^2$.}
\begin{ruledtabular} 
\begin{minipage}[b]{0.7\textwidth}
\begin{tabular}{lcccc}
Signal & $\epsilon{\cal{B}}\times 10^4$ & $N_{obs}$ & sig, $\sigma$ & $\Br$ $\times 10^4$\\

$ B \to \ddp K$    & 2.12$\pm$0.10  &24.1$\pm$6.1 & 6.4 &  1.27$\pm$ 0.31$_{-0.39}^{+0.22}$\\
$ B^+\to \ddp K^+$ & 3.62$\pm$0.14  &17.4$\pm$5.2 & 5.0 &  1.07$\pm$ 0.31$_{-0.33}^{+0.19}$\\
$ B^0\to \ddp K^0$ & 0.84$\pm$0.04  & 6.5$\pm$2.6 & 4.6 &  1.73$\pm$ 0.70$_{-0.53}^{+0.31}$\\
\end{tabular}
\end{minipage}
\end{ruledtabular}
\label{tab:bresult}
\end{center}
\end{table}
}

The MC-determined signal efficiency for the near-threshold peak is 
(1.87$\pm$0.05)\%. The contribution from non-resonant $B \to \ddst K$ events in the
near-threshold $M_{\ddp}$ region, calculated from a large 
sample ($\sim$ 60 times real data) is found to be 1.6 $\pm$ 0.2 events.

Because of the limited available phase space, the 
$M_{D^0\pi^0}/M_{\overline{D}{}^0\pi^0}$ 
distributions near the threshold show some 
clustering around $D^{*0}$ mass.
Thus it is not possible to separate the contributions
of $\ddst$ and $\ddp$ to the peak.

With a large sample of MC-simulated $\ddp$ events
generated in a narrow, near-threshold mass peak,
it is found that the reconstructed peak in the $M_{\ddp}$
distribution has a high-mass tail that is caused by
poorly reconstructed 
$\pi^0$'s. We are unable
to distinguish this high-mass tail component
in the data or in samples of MC signal plus background 
with sizes similar to the data. This tail,
if it exists, would produce a positive bias
on the peak mass measurement.  We account for this
possibility with asymmetric systematic errors on
the peak mass and efficiency of $^{+0.0}_{-1.7}$ MeV/$c^2$
and $^{+25.9}_{-~4.8}\%$, respectively. Including all systematics,
the observed $M_{\ddp}$ peak position is
$3875.4 \pm 0.7^{+0.4}_{-1.7} \pm 0.9$ MeV/$c^2$, where the
second error is mainly due to the calibration uncertainty
of the $\pi^0$ energy and the effects of a possible
high-mass tail as discussed above.  The third error is due to the
uncertainty in the world-average $D^0$ mass \cite{pdg2006}.

The systematic uncertainty on the $B \to \ddp K$  branching fraction for 
the near-threshold peak is obtained from the quadratic sum of the 
uncertainties due to (a) limited MC statistics (1.3\%), (b) subtraction 
of the $\bddstk$ contribution (5.1\%), (c) number of $\BB$ events 
($N_{B\overline{B}}$) (1.3\%), (d) PDG branching fraction of $D^0$ and $\ks$ 
(5.0\%), (e) track finding efficiencies (9.5\%), (f) 
$K / \pi$ identification uncertainties (7.0\%), estimated using 
    $D^{*-} \to \overline{D}{}^0 (\to K^+ \pi^-) \pi^-$ events,
(g) $\pi^0$ detection efficiency (7.0\%), estimated from a comparison of 
    $D^0 \to K^- \pi^+ \pi^0$ yields in data and MC,
(h) $\ks$ selection efficiency, estimated from a comparison of
    $D^0 \to \ks \pi^+ \pi^-$ yields in data and MC (2.1\%),  
(i) the ratio of $D^0$-mass window in data and MC (2.0\%),
(j) signal efficiency, calculated from the difference in $\de$ and $M_{\ddp}$ 
    fits (5.8\%),
(k) efficiency due to poorly reconstructed $\pi^0$ ($^{+~4.8}_{-25.9}$\%) and
(l) mass value (1.0\%).
The total uncertainty is estimated to be $^{+17.2}_{-30.7}$\%.

In summary, a near-threshold $\ddp$ invariant mass enhancement
is observed at 3875.4$\pm$0.7$^{+0.4}_{-1.7}\pm$0.9 MeV/$c^2$
in $B \to \ddp K$ decays. The significance of this enhancement is
6.4$\sigma$. 

The observed $\ddp$ mass is $2.0\sigma$ higher than 
the world-average value of the $X(3872)$ mass of 
3871.2 $\pm$ 0.5 MeV/$c^2$ while the branching fraction
of this threshold peak is 9.4$^{+3.6}_{-4.3}$ times larger
than $\Br(B^+ \to X(3872) K^+) \times \Br (X(3872) \to J/\psi \pi^+\pi^-)$
 \cite{pdg2006}.  Reference~\cite{properties1} ruled out all 
possible quantum states of $X(3872)$ except $J^{PC} = 1^{++}$ and 
$2^{++}$ while CDF finds that possible quantum number assignments are 
$1^{++}$ and $2^{-+}$ \cite{cdf2}.
If this near-threshold enhancement
is due to the $X(3872)$, the $J^{PC} = 1^{++}$ quantum number assignment
for the $X(3872)$ is favored, because the near-threshold decay 
$X(3872) \to \ddst/\ddp$ is expected to be strongly suppressed for $J=2$.

We thank the KEKB group for excellent operation of the
accelerator, the KEK cryogenics group for efficient solenoid
operations, and the KEK computer group and
the NII for valuable computing and Super-SINET network
support.  We acknowledge support from MEXT and JSPS (Japan);
ARC and DEST (Australia); NSFC and KIP of CAS (contract No.~10575109 and IHEP-U-503, China); DST (India); the BK21 program of MOEHRD, and the
CHEP SRC and BR (grant No. R01-2005-000-10089-0) programs of
KOSEF (Korea); KBN (contract No.~2P03B 01324, Poland); MIST
(Russia); ARRS (Slovenia);  SNSF (Switzerland); NSC and MOE
(Taiwan); and DOE (USA).


\begin{thebibliography}{99}
\bibitem{xdiscover}  Belle Collaboration, S. K. Choi {\it et al.}, Phys. Rev. 
                     Lett. {\bf 91}, 262001 (2003).
\bibitem{x3872oth}   CDF Collaboration, D.~Acosta {\it et al.}, 
                     Phys. Rev. Lett. {\bf 93}, 072001 (2004),   
                     D0 Collaboration, V.M.~Abazov {\it et al.}, 
                     Phys. Rev. Lett. {\bf 93}, 162002 (2004),  
                     BaBar Collaboration, B.~Aubert {\it et al.}, 
                     Phys. Rev. {\bf D71}, 071103 (2005).
\bibitem{properties} Belle Collaboration, K. Abe {\it et al.}, 
                     hep-ex/0408116. 
\bibitem{hybrid}     E. F. Close and S. Godfrey, Phys. Lett. {\bf B574} 210
                     (2003).
\bibitem{deuson_ref} N. T$\ddot{\rm o}$rnqvist, 
                     Phys. Lett. {\bf B590} 209 (2004).
\bibitem{braaten}    E. Braaten, M. Kusunoki and S. Nussinov, Phys. Rev. Lett. 
                     {\bf 93}, 162001 (2004),  E. Braaten and M. Kusunoki, 
                     Phys. Rev. {\bf D69}, 074005 (2004).
\bibitem{nullres}    Belle Collaboration, R.~Chistov {\it et al.}, 
                     Phys. Rev. Lett. {\bf 93}, 051803 (2004).

\bibitem{voloshin}  M. B. Voloshin, Phys. Lett. {\bf B579}, 316 (2003).
\bibitem{belledet}  Belle Collaboration, A. Abashian {\it et al.},
                    Nucl. Instr. Meth. {\bf A479}, 117 (2002).
\bibitem{kekb}      S. Kurokawa and E. Kikutani,
                    Nucl. Instr. Meth. {\bf A499}, 1 (2003).
\bibitem{ksref}     Belle Collaboration, M. Nakao {\it et al.}, 
                    Phys. Rev. {\bf D69}, 112001 (2004).
\bibitem{foxwol}    G. C. Fox and S. Wolfram,
                    Phys. Rev. Lett. {\bf 41}, 1581 (1978).
\bibitem{bbarbtocck} BaBar Collaboration, B.~Aubert {\it et al.},
                     Phys. Rev. {\bf D68}, 092001 (2003).
\bibitem{pdg2006}   W.-M. Yao (Particle Data Group), private communication.
\bibitem{properties1} Belle Collaboration, S. K. Choi {\it et al.}, hep-ex/0505038.
\bibitem{cdf2}       CDF Collaboration, A.~Abulencia {\it et al.}, 
                     Phys. Rev. Lett. {\bf 96}, 102002 (2006).

\end{thebibliography}
\end{document}